\documentclass[conference]{IEEEtran}
\IEEEoverridecommandlockouts
\IEEEoverridecommandlockouts
\IEEEpubid{\makebox[\columnwidth]{MOCAST 2025 -- Paper Preprint  \copyright2025 IEEE \hfill}
\hspace{\columnsep}\makebox[\columnwidth]{ }}

\usepackage{amsmath,amssymb,amsfonts}
\usepackage{algorithmic}
\usepackage{graphicx}
\usepackage{textcomp}
\usepackage{xcolor}
\usepackage{booktabs}
\usepackage{array}
\usepackage{float}
\usepackage{orcidlink}

\usepackage{amsmath}
\usepackage{siunitx}
\usepackage{makecell}
\newcolumntype{N}{>{\centering\arraybackslash}m{1.0in}}
\newcolumntype{M}[1]{>{\centering\arraybackslash}m{#1}}

\newcommand{\squeezeCaption}{\vspace{0mm}}
\newcommand{\squeezeText}{\vspace{0mm}}
\def\BibTeX{{\rm B\kern-.05em{\sc i\kern-.025em b}\kern-.08em
    T\kern-.1667em\lower.7ex\hbox{E}\kern-.125emX}}
\begin{document}

\title{VUSA: Virtually Upscaled Systolic Array Architecture to Exploit Unstructured Sparsity in AI Acceleration\vspace{0mm}}




\author{\IEEEauthorblockN{ Shereef Helal}
\IEEEauthorblockA{\textit{NXP Semiconductors} \\
Munich, Germany \vspace{0mm}\\
shereef.helal@nxp.com
}
\and
\IEEEauthorblockN{Alberto Garcia-Ortiz}
\IEEEauthorblockA{\textit{University of Bremen} \\
Bremen, Germany \vspace{0mm}\\
agarcia@item.uni-bremen.de
}
\and
\IEEEauthorblockN{Lennart Bamberg}
\IEEEauthorblockA{\textit{NXP Semiconductors} \\
Hamburg, Germany \vspace{0mm}\\
lennart.bamberg@nxp.com
}
}

\maketitle
\IEEEpubidadjcol  

\begin{abstract}
Leveraging high degrees of unstructured sparsity is a promising approach to enhance the efficiency of deep neural network (DNN) accelerators---particularly important for emerging Edge-AI applications. We introduce VUSA, a systolic-array architecture that virtually grows based on the present sparsity to perform larger matrix multiplications with the same number of physical multiply-accumulate (MAC) units. The proposed architecture achieves saving by 37\% and 68\% in area and power efficiency, respectively, at the same peak-performance, compared to a baseline systolic array architecture in a commercial 16-nm technology. Still, the proposed architecture supports acceleration for any DNN with any sparsity---even no sparsity at all. Thus, the proposed architecture is application-independent, making it viable for general-purpose AI acceleration.
\end{abstract}


\section{Introduction}
Over recent years, Artificial Intelligence (AI) has emerged as a revolutionary new technology, spreading across different industries and enhancing various aspects of our daily lives.
The deployment of AI is not only confined to powerful data-center machines, but is increasingly demanded in resource-constrained embedded devices, a concept known as Edge AI.
Deep Neural Network (DNN) architectures are the backbone of state-of-the-art AI applications to perform numerous tasks, such as image processing, speech recognition, natural language processing (NLP), and more~\cite{cnn}. However, DNNs have high computational demands, posing a significant challenge when deploying them in real-world applications. This challenge is even greater for Edge AI applications in resource-constrained environments.




Domain-specific hardware accelerators are widely used for Edge AI applications to improve the efficiency of DNN inference~\cite{tpu,spots,sense}. Due to its scalability and programmability, a systolic array is one of the most widely used architectures for DNN acceleration. The scalability and programmability come from a regular and simple structure that allows efficient matrix-by-matrix multiplications. Such matrix multiplications account for the largest fraction of the computational complexity of modern AI applications.
However, the regularity of the systolic array also brings drawbacks such as poor utilization for large array dimensions (i.e., large number of multiply-accumulate (MAC) blocks) and no exploitation of unstructured sparsity. 

Weight sparsity refers to the number of zero-valued parameters in a DNN. The computations associated with those parameters can be skipped without affecting the final results, thereby improving power efficiency, latency, and memory usage.
Most accelerators can only exploit structured sparsity, which limits the amount of usable sparsity to 50\% and has poor accuracy. In contrast, modern DNN applications can be trained to achieve much higher sparsity levels—exceeding 90\% without any loss in accuracy when sparsity is unconstrained (i.e., unstructured)~\cite{deepcompression}.
Thus, extending regular and scalable accelerator architectures, such as systolic arrays, to support this form of sparsity is a topic of great research interest in both academia~\cite{eie} and industry~\cite{bamberg2023synapse,bamberg2024exploiting}.


This work extends the standard weight-stationary systolic array architecture to exploit unstructured sparsity as much as possible while maintaining high efficiency for dense (i.e., non-sparse) execution. Specifically, we modify the architecture so that the systolic array can virtually expand to a larger size in the presence of weight sparsity. This expansion increases computational speed to match that of a more hardware-expensive array. Nevertheless, the simplicity of the hardware architecture and programmability is preserved, maintaining the key advantages of systolic arrays in terms of scalability and general-purpose applicability. Experimental results in a commercial 16-nm technology demonstrate that the proposed technique improves the area and power consumption by 37\% and 68\%, respectively, while maintaining the same peak performance.


In the proposed architecture, the maximum amount of exploitable weight sparsity is defined by an integer architectural parameter --the maximum virtual growth. While the standard systolic array size is determined by the number of rows and columns, the proposed \textit{virtually upscaled systolic array} (VUSA) is characterized by the number of rows, columns, and virtual columns. This approach preserves the generic and flexible nature of systolic arrays while expanding the design space to better exploit sparsity. To the best of the authors' knowledge, this is the first systolic array architecture that introduces a new dimensionality to effectively leverage unstructured weight sparsity in a weight-stationary systolic array without compromising scalability or programmability.


The rest of the paper is structured as follows. Section~\ref{sec:bg} outlines the background. Subsequently, in Section~\ref{sec:concept}, we outline the architecture of the proposed VUSA.  A quantitative, theoretical analysis of the expected gain of the proposed technique is presented in Section~\ref{sec:theo_anal}.
Section~\ref{sec:results} includes experimental results for a commercial 16-nm technology and real-life DNN applications. Finally, a comparison to related work and a conclusion are drawn.


\section{Background}
\label{sec:bg}

\subsection{Systolic Arrays}
The structure of a systolic array is shown in Figure 1.
Systolic arrays are used to accelerate tensor operations by decomposing them into matrix-by-matrix multiplications (in combination with data arrangements by other components). The architecture is made up of multiple processing elements (PEs), performing MAC operations, between which data automatically flows from the left to the right and top to the bottom. Such a data-driven architecture heavily reduces the memory accesses per compute, enabling scaling to compute capabilities in the high TOPS regime~\cite{why_systolic}.
Common data flows in systolic arrays are: Weight Stationary (WS), Input Stationary (IS), and Output Stationary (OS). The name of the dataflow indicates which parameter is stationary to the PEs during a processing sequence, while the other two are flowing from left to right and top to bottom~\cite{scalesim}. This work focuses on WS dataflow as it enables to exploit unstructured weight sparsity while keeping the flow of data regular.

\begin{figure}[t]
     \centering
    \includegraphics[trim=50 170 70 180,clip,width=0.48\textwidth]{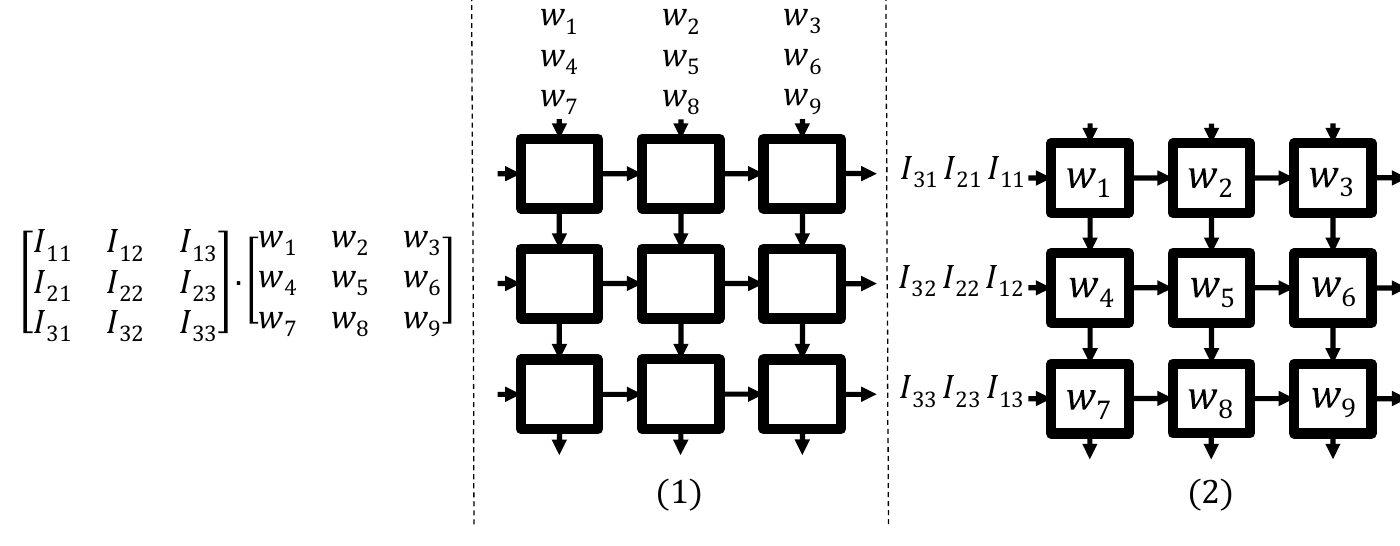}
    \squeezeCaption
    \caption{Systolic array with weight-stationary data flow.}
    \label{fig:2_WS}
    \squeezeText
\end{figure}

An example WS matrix multiplication is illustrated in {Figure~\ref{fig:2_WS}}.
First, the weights are loaded into the PEs. The weights remain stationary to the respective PEs until the computations are done. Afterwards, the inputs are fed to the PEs from the left and multiplied by the weights to generate partial sums. Each processing element adds the partial sum to the accumulator value from the top input. To the right output the left input is forwarded as is, while to the bottom the updated accumulator value is passed. Multiplications of matrices larger than the systolic array size are decomposed into multiple smaller pieces that fit the array size. For this purpose, the resulting accumulator values (i.e., the results flowing out at the bottom) from the previous step are used as the starting accumulator values (i.e., feed to the array at the top). Matrices smaller than the systolic array size lead to an under utilization of the systolic array.

\subsection{Unstructured Weight Sparsity}
The term sparsity describes the high degree of zero in neural network activations/data and weights.
Pruning weights is the primary source of high sparsity; A technique that removes low-magnitude weights with the smallest impact on the model output. If pruning is done iteratively during training, the accuracy can be preserved or even enhanced (reduced over-fitting), while achieving sparsity factors above 90\%~\cite{deepcompression}. However, dense accelerators such as standard systolic arrays cannot work with the compressed, pruned weights due to the absence of regular data flow. Thus, processing unstructured sparse weight tensors on such architectures, implies a large degree of zero-values in the weight matrices~\cite{spots}. In this case, the effective operation of the large fraction of MAC blocks with a stationary zero-weight becomes a pass-through of both accumulator and input value (i.e., a NoOP).


\section{Proposed Architecture}
\label{sec:concept}
Our goal is to extend the systolic-architecture to be capable of working with any weight matrix in which the total number of non-zero weights does not exceed the number of MAC units. In this case, each non-zero weight is loaded to one MAC, ignoring zero weights since they do not contribute to the final result. 
The challenge that this destroys the regularity of the data-flow of inputs and activations is addressed by extending the architecture as follows.
\subsection{Processing Sparse Weights in PEs}
\begin{figure}[]
    \centering
    \includegraphics[trim=130 140 140 120,clip,width=0.35\textwidth]{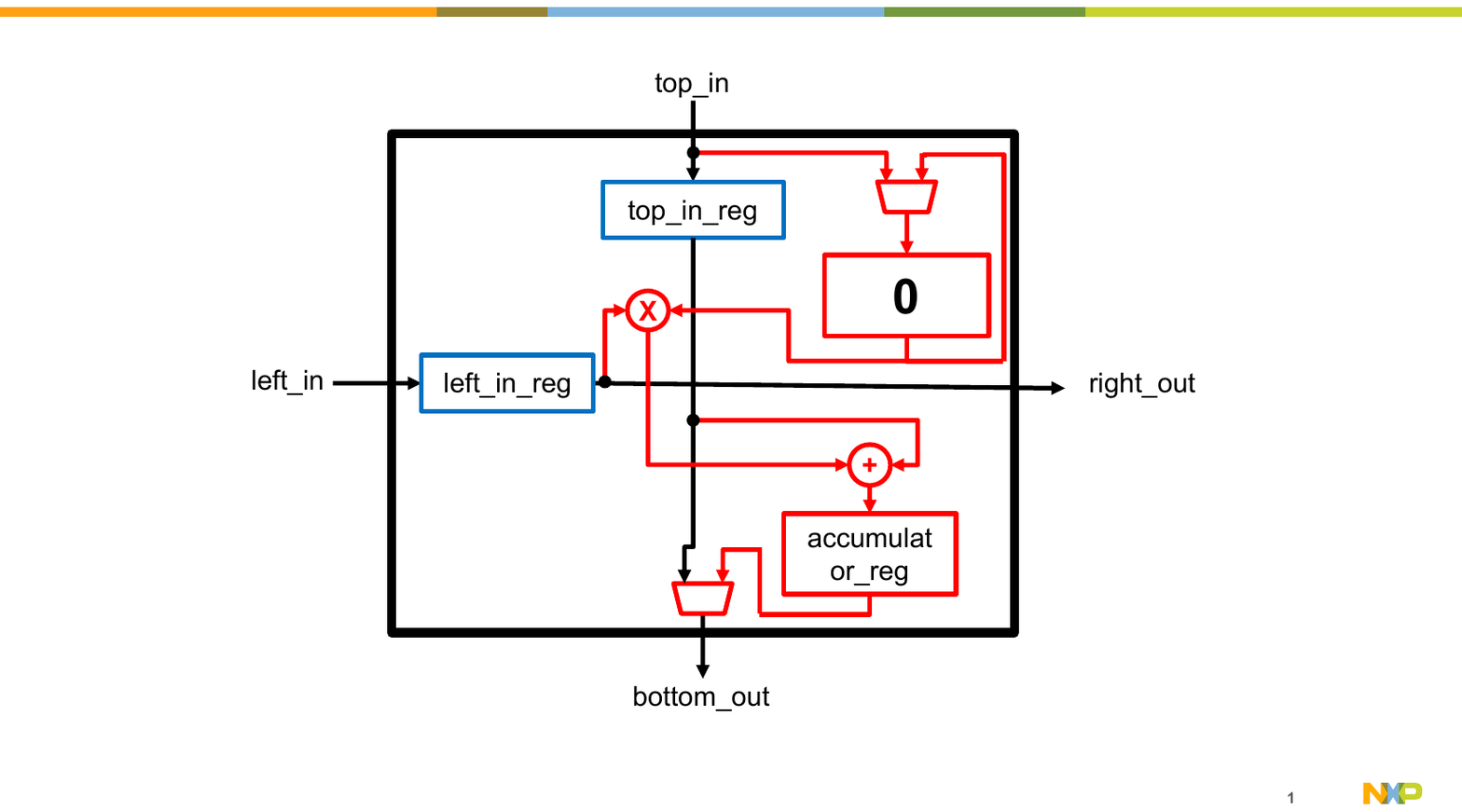}
    \squeezeCaption
    \caption[Removable logic inside PE]{Example architecture of a PE. Parts marked in red can be removed when the stationary weight is zero.}
    \label{fig:4_pe_ws_simple}
    \squeezeText
\end{figure}

The operation of a weight stationary PE is distinguished here in two different operating states:
\begin{enumerate}
    \item \textbf{Non-Sparse: stationary weight is non-zero:} In this case, the full logic inside the PE depicted in {Figure~\ref{fig:4_pe_ws_simple}} is required for correct processing.
    \item \textbf{Sparse: stationary weight is zero}: In this case, no data-path blocks are required. Only the pipeline stages for the cycle-by-cycle data-flow are required. Figure~\ref{fig:4_pe_ws_simple} highlights in different colors the parts that can be removed for sparse processing (red) versus what must be kept to keep the data-flow regular (blue).
\end{enumerate}

\subsection{Separating Processing from Data-Flow Elements}
\begin{figure}[]
    \centering
    \includegraphics[trim=65 170 105 150,clip,width=0.45\textwidth]{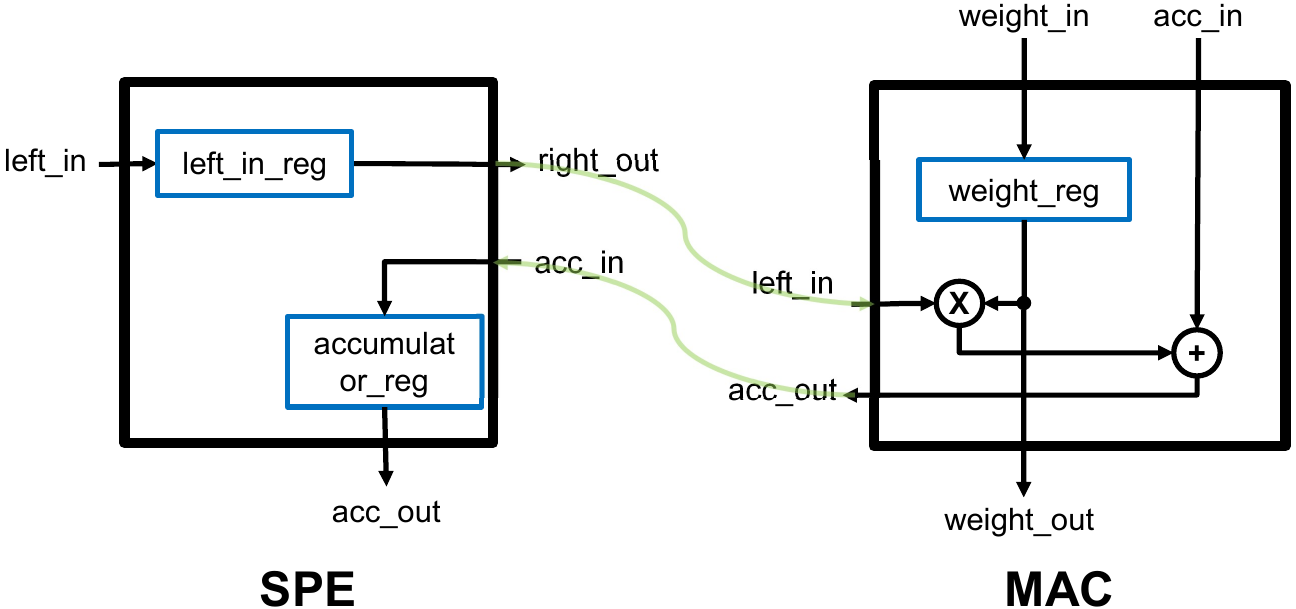}
    \squeezeCaption
    \caption{Decomposing the PE to a data-flow and a compute block.}
    \label{fig:4_spe_mac}
    \squeezeText
\end{figure}

Inspired by the fact that no compute element is actually needed when the weight is zero, we decompose the PE into two components: the MAC unit (i.e., processing components), and the pipeline stages (data-flow components). 
The resulting two blocks are shown in {Figure~\ref{fig:4_spe_mac}}.
\begin{itemize}
    \item \textbf{Sparse Processing Element (SPE):} A block containing only pipeline registers, required for sparse as well as non-sparse processing
    \item \textbf{MAC unit:} A block that can be optionally connected to an SPE, required only for non-zero weights. 
\end{itemize}

\subsection{Virtual Upscaled Systolic Array (VUSA)}
\label{sec:ssa_concept}
As MAC units are not required for zero weights, they can be instantiated in a lower dimension than the maximum virtual systolic array size. Based on that, a sparse $N\times M$ systolic array with $N\times A$ MAC units is proposed, where $N$ is the number of rows, $M$ is the number of columns, and $A$ is the number of MAC units per row. Thus, each row has $M$ SPEs but only $A$ number of MAC units where $1\leq A \leq M$.

If a weight is zero, then only an SPE is sufficient in that position. Otherwise, a MAC unit is connected to the SPE in that position for the needed arithmetic computation. If the number of non-zero weights per row is less than or equal to the number of available MAC units per row ($A$), then the sparse systolic array virtually grows to an $N\times M$ systolic array. Otherwise, a smaller window of weights is selected until the sparsity condition is satisfied, starting with an $N\times (M-1)$ window, then $N\times (M-2)$, and so on down to $N\times A$, at which the conditions are guaranteed to be satisfied. Thus, still any matrix-matrix multiplication can still be mapped--even in the absence of sparsity. In detail, no sparsity just implies slower processing as we need to break down larger matrix-by-matrix multiplications into sub-blocks of size $N\times A$.

\begin{figure}[]
    \centering
    \includegraphics[trim=40 200 40 170,clip,width=0.35\textwidth]{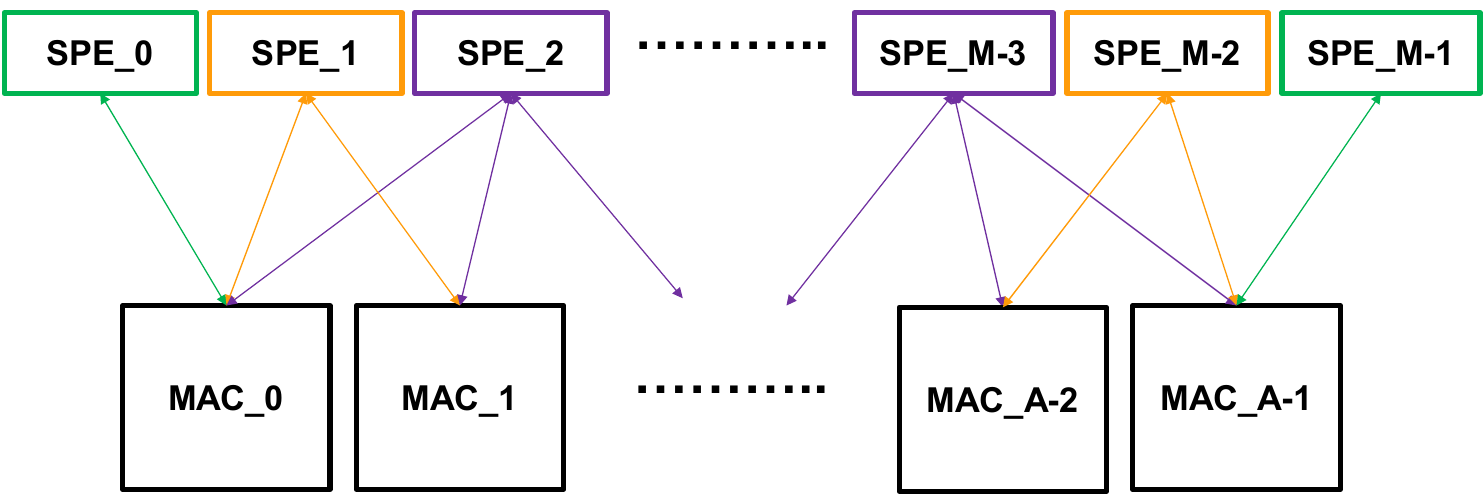}
    \squeezeCaption
    \caption{Single row of the proposed  VUSA architecture.}
    \label{fig:4_sparse_row}
    \vspace{-3mm}
\end{figure}

A conceptual illustration of a single row of the VUSA is shown in Figure~\ref{fig:4_sparse_row}. Each MAC unit has a configurable connection to multiple SPEs. Depending on the positions of non-zero weights in each row, the MAC units are connected to SPEs where computations involving non-zero weights are required. The connections to different MAC units can be dynamically configured such that changing positions of non-zero weights can be handled per job. Still, an all-to-all connection between SPEs and MAC units is not needed. Each MAC unit only needs to be connected to one out of $M-A+1$ directly adjacent SPEs to handle all potential zero-distributions. 
Let us label the SPEs with indexes $i \in [0,...,M-1]$ and the MACs with $j \in [0,...,A-1]$. Consider now $MAC$ number $r$, associated with SPE number $t$, MAC $r$ can be mapped only to SPEs $[r,...,r+M-A]$.
In other words, an input-output shifter of up to $M-A$ positions per MAC in one direction is sufficient. Thereby, wiring is minimized which in turn reduces negative impacts on timing and power. In fact, experimental results for a commercial 16-nm technology (cf. Section~\ref{sec:results}) show that even for $M\gg A$ the critical timing path and the power bottleneck is the MAC unit and not the MAC-to-SPE multiplexing.


\begin{figure}[]
    \centering
    \includegraphics[trim=30 190 30 177,clip,width=0.35\textwidth]{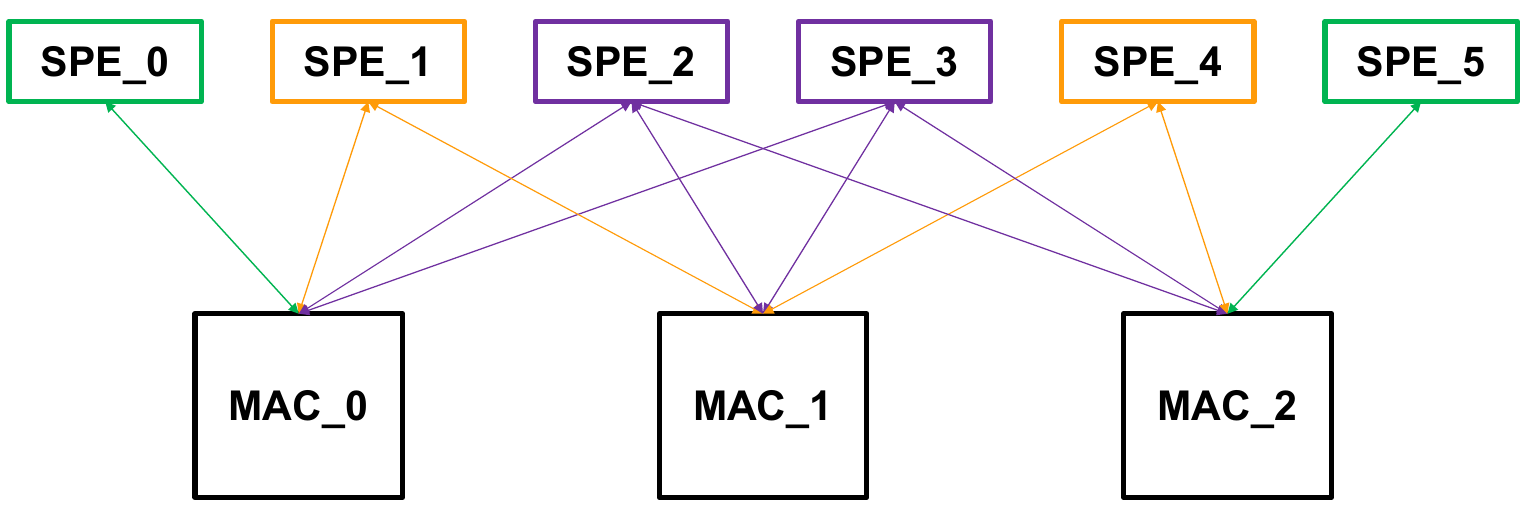}
    \squeezeCaption
    \caption{Row instance of the proposed VUSA for $M=6$ and $A=3$.}
    \label{fig:5_mux_3-6}
    \squeezeText
\end{figure}

For example, applying this design to a sparse row with $M=6$ and $A=3$ requires to connect each MAC to four SPEs. This is illustrated in {Figure~\ref{fig:5_mux_3-6}}.

This proposed architecture can handle unstructured sparsity because the position on non-zero weights is not restricted. It achieves direct savings in resources by using the same number of MAC units to conduct matrix multiplications of a higher dimension. Since MAC units are not connected in positions where weights are zero, a natural multiplier gating~\cite{cnvlutin,eyeriss} for sparse weights is achieved. In this case, the gating applies not only to the multiplier but to the full MAC unit. The sparse row is the building block of VUSA which can be stacked vertically or horizontally to implement VUSAs with arbitrary dimensions.

\section{Theoretical Analysis}\label{sec:theo_anal}
In this section, we statistically assess the theoretical performance gains of the proposed architecture as a function of the sparsity. Truly unstructured sparsity implies that there is no correlation of sparsity between the individual weight positions. Thus, the sparsity probability of each weight is the same with no extra cross-correlation (i.e., inter-dependencies between the sparsity of different weights). In other words, the number of zero-terms follow a Binomial distribution.

The maximum gain of the VUSA is achieved when the array dimension virtually grows to $N\times M$. Under the above assumption of no cross-correlation, the probability of this occurring depends only on the probability that a weight is non-sparse, denoted here as $P_1$. For a VUSA, defined by tuple $(N,M,A)$, the probability that the VUSA virtually grows to a standard $N\times M$ systolic array, $P_{sparse \rightarrow standard}$, can be derived as follows:
\begin{enumerate}
    \setcounter{enumi}{0}
    \item The gain probability of a single row ($P_{gain\_row}$) is computed which is the probability that the number of non-zero weights in the set of $M$ is less than or equal to the number of available MAC units per row, $A$. This can be calculated by summing up the probabilities that the number of non-zero weights is exactly one of the numbers in the range $[0{:}A]$. The probability of getting exactly $i$ number of non-zero weights is denoted as $P_{i,ones}$. Based on that, $P_{gain\_row}$ can be calculated as:
\end{enumerate}

\begin{equation}\label{eqn:1}
P_{gain\_row} = \sum\limits_{i=0}^A P_{i,ones}
\end{equation}
\begin{enumerate}
    \setcounter{enumi}{1}
    \item The gain of the VUSA is achieved when the sparsity conditions are met for all rows. Hence, $P_{sparse \rightarrow standard}$ can be derived as:
\end{enumerate}
\begin{equation}
P_{sparse \rightarrow standard} = {(P_{gain\_row})}^N = \left(\sum\limits_{i=0}^A P_{i,ones}\right)^N
\label{eqn:2}
\end{equation}

\begin{enumerate}
    \setcounter{enumi}{2}
    \item $P_{i,ones}$ can be calculated as:
\end{enumerate}
\begin{equation}
P_{i,ones} = \binom{M}{i}{P_1}^i(1-P_1)^{M-i}
\label{eqn:3}
\end{equation}

\begin{enumerate}
    \setcounter{enumi}{3}
    \item From {Equations~\ref{eqn:2} and~\ref{eqn:3}}, the formula for $P_{sparse \rightarrow standard}$ becomes:
\end{enumerate}
\begin{equation}\label{eqn:4}
P_{sparse \rightarrow standard} = \left(\sum\limits_{i=0}^A \binom{M}{i}{P_1}^i(1-P_1)^{M-i}\right)^N
\end{equation}
If desired, a simpler analytical expression can be drawn using the regularized incomplete beta function, the CDF of the binomial, but the explicit result as a sum is more intuitive.


Analogously, one can express the probability to have a virtual growth in the array size to another $M'$, smaller than the theoretical maximum $M$ defined by the hardware configuration, but larger than $A$. The probability to map a given problem to a $N\times A$ array (no virtual extension) is 1, as this is always possible, independent of the weights.

\begin{figure}[]
    \centering
    \includegraphics[trim=120 150 130 180,clip,width=0.45\textwidth]{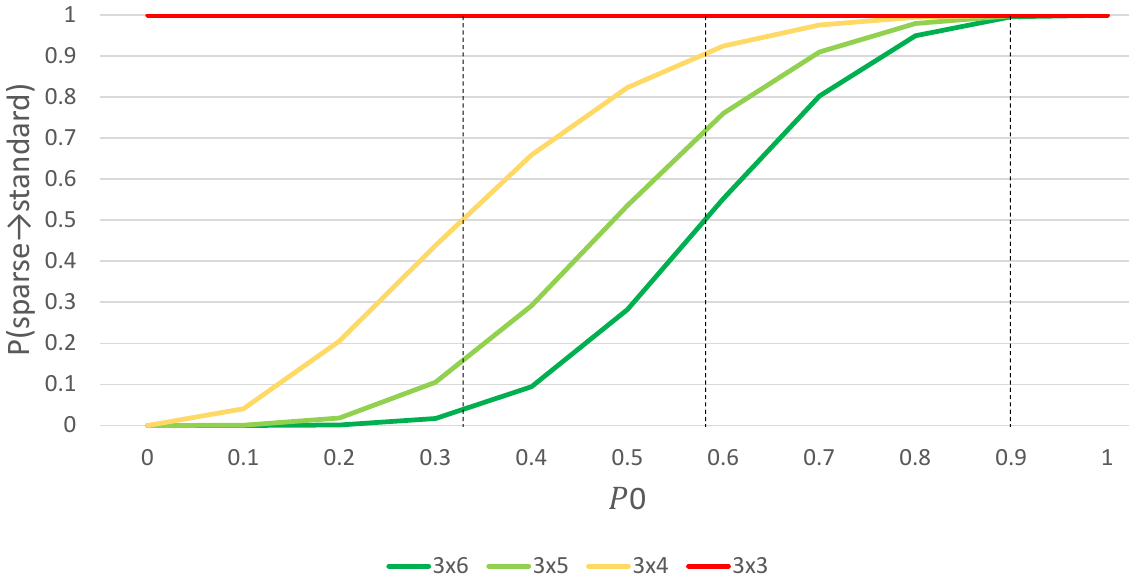}
    \squeezeCaption
    \caption{Growth probability of VUSA with $N=3$, $M=6$, and $A=3$}
    \label{fig:6_p_success}
    \squeezeText
\end{figure}

By applying the derived mathematical model to a VUSA with $N=3$, $M=6$, and $A=3$, the probability to be able to virtually grow to $3\times 6$, $3\times 5$, $3\times 4$ is calculated based on the sparsity rate, $P_0 = 1-P_1$. The results are shown in {Figure~\ref{fig:6_p_success}}.
It can be observed that when sparsity is higher than $90\%$, the probability to virtually grow to $3\times 6$ (i.e., $2\times$ speed-up) is close to 1. Even for sparsity rates as low as 60\%, the success rate for the maximum gain is above 50\%. Even at dramatically low sparsity of around $30\%$, the success rate to grow to a $3\times 4$ array (i.e., $\tfrac{4}{3}\times$ speed-up) is still above 50\%.
Only in the theoretical worst case of no exploitable sparsity ($3\times 3$ window), there is no performance gain. On a real full AI application, composed of many different matrix multiplication, the VUSA performs with a combination of all virtual growth possibilities based on the weight sparsity of the current computation. This behavior is analyzed by means of experimental results next.

\section{Experimental Results}
\label{sec:results}
In this section, a VUSA with $N=3$, $A=3$, $M=6$ (i.e., 9 MAC units that can virtually grow to 18) is evaluated against varying systolic array sizes for real benchmark applications.
The standard systolic array dimensions to be investigated are: $3\times 3$, $3\times 4$, $3\times 5$, and $3\times 6$.

\subsection{Physical-Design}
The different architectures are physical-aware (to consider interconnects) synthesized in a commercial 16-nm technology at the same target frequency of 1GHz.
{Table~\ref{tab:6_all_designs_asic}} shows the resulting silicon area and power consumption for all designs normalized to the values of the proposed VUSA architecture. 
\begin{table}[]
    \centering
        \caption{Synthesis results in a commercial 16-nm technology at \SI{1}{\giga\hertz}}
        \scriptsize
    \begin{tabular}{cccc}
         \toprule
         Design& \#MACs& Area (Normalized)&  Power (Normalized)\\
         \midrule
         Standard 3$\times$3& 9& 0.69&  0.86\\
         
         Standard 3$\times$4& 12& 0.91&  1.15\\
         
         Standard 3$\times$5& 15& 1.14&  1.41\\
         
         Standard 3$\times$6& 18& 1.37&  1.68\\
         \midrule
         \textbf{VUSA 3$\times$6}& \textbf{9}&  \textbf{1}&  \textbf{1}\\
         \bottomrule
    \end{tabular}
    \squeezeText

    \label{tab:6_all_designs_asic}
\end{table}
The area footprints of standard $3\times 6$ and $3\times 5$ systolic arrays are $37\%$ and $14\%$,  larger than the proposed  VUSA architecture, respectively.
On the other end, the areas of standard $3\times 4$ and $3\times 3$ arrays are $9\%$ and $31\%$ smaller, respectively. In terms of power consumption, a standard $3\times 6$, $3\times 5$, and $3\times 4$ array are $68\%$, $41\%$, and $15\%$, more costly than the proposed solution, respectively. A standard $3\times 3$ array has a $14\%$ lower power consumption.

Overall, there is a clear advantage in both area and power for the $3\times 6$ and $3\times 5$ cases. For the $3\times 4$ case, there are gains in power but with a small area overhead. Compared to a standard $3\times 3$ array, there are area and power overheads. Thus,  to be an efficient solution on a combined power-performance-area metric, our proposed technique must dominantly work at least as $3\times 4$ virtual systolic array for a given application.

\subsection{Effect of Sparsity on Performance}
\label{sec:sparse_effect}
The number of zero weights and their positions in the different layers of a given application determines the number of jobs to be executed on the VUSA structure. If the weights of a given load are almost all zeros, the VUSA virtually grows to a $3\times 6$ array, implying a significant gain over the standard architecture. If on the other hand all weights are non-zero, then the VUSA acts only as a $3\times 3$ and thus performs worse than a standard systolic array due to the power and area overhead outlined above.

\begin{figure}[]
    \centering
    \includegraphics[trim=190 220 170 190,clip,width=0.35\textwidth]{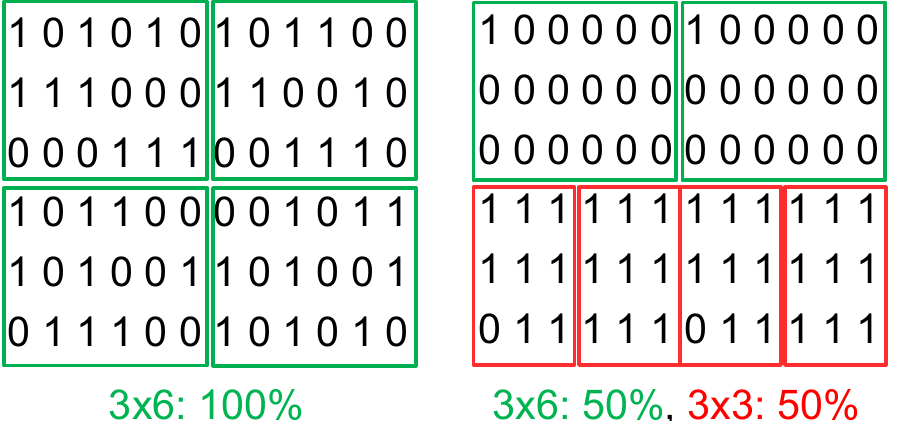}
    
    \caption{Example $50\%$ sparsity distribution}
    \label{fig:6_sparse_example}
    \squeezeText
\end{figure}

Not only the number of zeros but also their distribution is relevant. Consider a load with $50\%$ sparsity, as depicted in {Figure~\ref{fig:6_sparse_example}}. If the zeros are distributed approximately evenly, then the VUSA can process the entire problem with a $3\times 6$ virtual array. On the other hand, if the zeros are correlated in their positions, then the VUSA processes only $50\%$ of the jobs with a $3\times 6$ window,  while the other $50\%$ is processed with a $3\times 3$ window. On a real application with unstructured sparsity, a combination of all cases typically occurs. 

\subsection{Methodology to Analyze Full DNN Models}
\label{sec:method}
The evaluation of full applications is performed in two steps: first, a cycle-accurate simulation is performed using SCALE-Sim~\cite{scalesim} to obtain the number of required cycles to run DNN models on standard systolic arrays of dimensions from $3\times 3$ to $3\times 6$. The second step is obtaining pre-trained models and using the values of the pre-trained weights to determine the sparsity conditions of the weights (number of zeros and their distribution). Based on the weight filters, the number of different windows of sizes from $3\times 3$ to $3\times 6$ required by the VUSA is determined, which can be used to calculate the percentages of the VUSA performing as $3\times 6$, $3\times 5$, $3\times 4$, and $3\times 3$ systolic arrays. Finally, those percentages combined with the simulation results are used to calculate the performance of the VUSA compared to the reference architecture.

\subsection{Efficiency on real DNN Models}
After computing the number of cycles to execute the target model on different designs following the described methodology in {Section~\ref{sec:method}}, the execution time and performance is calculated for standard Edge-AI benchmarks. Afterwards, the area efficiency (performance per area unit), power efficiency (performance per power unit), and total energy consumption are calculated. For the following experiments, pre-trained models with unstructured sparsity from SparseZoo~\cite{sparsezoo} are deployed.

\subsubsection{ResNet}
The first evaluated model is ResNet-18~\cite{resnet} pruned by $85\%$. The evaluation results are depicted in {Table~\ref{tab:6_performance_resnet}}.
\begin{table}[]
    \centering
        \caption{Performance on ResNet-18 pruned at $85\%$}
        
        \scriptsize
    \begin{tabular}{ccccc}
         \toprule
         Design&  \thead{Load \\ split}&  \thead{Number of \\ cycles}& \thead{Time at \\ \SI{1}{\giga\hertz (\si{\milli\s})}} &\thead{Performance \\ (\si{\giga OP}/\si{s})}\\
         \midrule
         Standard 3$\times$3&   5.13\%&  \num{1.75E+08}& 174.70&	8.84 \\

         Standard 3$\times$4&   4.13\%&  \num{1.30E+08}& 130.03&	11.88 \\

         Standard 3$\times$5&   3.89\%&  \num{1.06E+08}& 106.20&	14.55 \\

         Standard 3$\times$6&  86.85\%&  \num{8.98E+07}&  89.81&	17.21 \\
         \midrule
         \textbf{VUSA 3$\times$6}&    & \textbf{\num{9.65E+07}}& \textbf{96.46}& \textbf{16.02} \\
         \bottomrule
    \end{tabular}

    \vspace{3pt}
    
    \begin{tabular}{cccc}
         \toprule
         Design&  \thead{Performance/Area \\ (Normalized)}& \thead{Performance/Power \\ (Normalized)}& \thead{Energy \\ (Normalized)}\\
         \midrule
         Standard 3$\times$3&  1.02& 1.00& 1.00 \\

         Standard 3$\times$4&  1.03& 1.01& 0.99 \\

         Standard 3$\times$5&  1.01& 1.00& 1.00 \\

         Standard 3$\times$6&    1&     1&    1 \\
         \midrule
         \textbf{VUSA 3x6}&   \textbf{1.27}& \textbf{1.56}& \textbf{0.64} \\
         \bottomrule
    \end{tabular}
    \vspace{0mm}

    \label{tab:6_performance_resnet}
\end{table}
Overall, VUSA performs as a standard $3\times 6$ for 86.85\% of the model load. Thus, it achieves a 10\% higher performance than a standard $3\times 5$ array with an additional 14\% area gain and 41\% power savings, showing a clear advantage of the proposed technique. In general, the proposed technique outperforms the standard array in any shape in all efficiency metrics by about 25\% in area and 55\% in power.

\subsubsection{MobileNetV1}
The second model used for evaluation is MobileNetV1~\cite{mobilenet} pruned at $75\%$. MobileNets are designed for resource-constrained edge devices and thus are particularly relevant. Also, they are harder to prune, hence the sparsity rate is lower. The MobileNet results are depicted in {Table~\ref{tab:6_performance_mobilenet}}.

\begin{table}[]
    \centering
        \caption{Results on MobileNetV1 pruned at $75\%$}
\scriptsize
    \begin{tabular}{ccccc}
         \toprule
         Design&  \thead{Load \\ split}&  \thead{Number of \\ cycles}& \thead{Time at \\ \SI{1}{\giga\hertz (\si{\milli\s})}} &\thead{Performance \\ (\si{\giga OP}/\si{s})}\\
         \midrule
         Standard 3$\times$3&  12.38\%&  \num{6.95E+07}& 69.50&	8.19 \\

         Standard 3$\times$4&  11.44\%&  \num{5.34E+07}& 53.43&	10.65 \\
 
         Standard 3$\times$5&   7.54\%&  \num{4.44E+07}& 44.39&	12.82 \\

         Standard 3$\times$6&  68.64\%&  \num{3.82E+07}& 38.16&	14.91 \\
         \midrule
         \textbf{VUSA 3$\times$6}&         &    \textbf{\num{4.43E+07}}& \textbf{44.25}&	\textbf{12.86} \\
         \bottomrule
    \end{tabular}

    \vspace{3pt}
    
    \begin{tabular}{cccc}
         \toprule
         Design&  \thead{Performance/Area \\ (Normalized)}& \thead{Performance/Power \\ (Normalized)}& \thead{Energy \\ (Normalized)}\\
         \midrule
         Standard 3$\times$3&  1.09& 1.07& 0.94 \\

         Standard 3$\times$4&  1.07& 1.04& 0.96 \\

         Standard 3$\times$5&  1.03& 1.02& 0.98 \\

         Standard 3$\times$6&    1&     1&    1 \\
         \midrule
         \textbf{VUSA 3$\times$6}&    \textbf{1.18}& \textbf{1.45}& \textbf{0.69} \\
         \bottomrule
    \end{tabular}
    \squeezeText

    \label{tab:6_performance_mobilenet}
\end{table}
Due to the lower sparsity, here the VUSA performs as a standard $3\times 6$ for 68.85\% of the model load. Thus, it achieves a similar performance as a standard $3\times 5$ array at an additional 14\% area gain and 41\% power savings. This shows the advantage of the proposed technique also for highly optimized Edge-AI model architectures with less unstructured sparsity. Again, the proposed technique outperforms the standard array in any shape in all efficiency metrics---here, by about 10--15\% in area and 40\% in power.

\subsection{Pruning Effect on Efficiency}
The ratio of jobs the VUSA computes as a $3\times 6$, $3\times 5$, $3\times 4$, and $3\times 3$ array determine the gains of the proposed technique. The weight sparsity is the major parameter defining these ratios. In this subsection, the area and power efficiencies are calculated for models pruned at different percentages to estimate the turning point at which using the VUSA is no longer beneficial over a standard systolic array.

\begin{figure}[]
    \centering
    \includegraphics[trim=40 210 300 245,clip,width=0.48\textwidth]{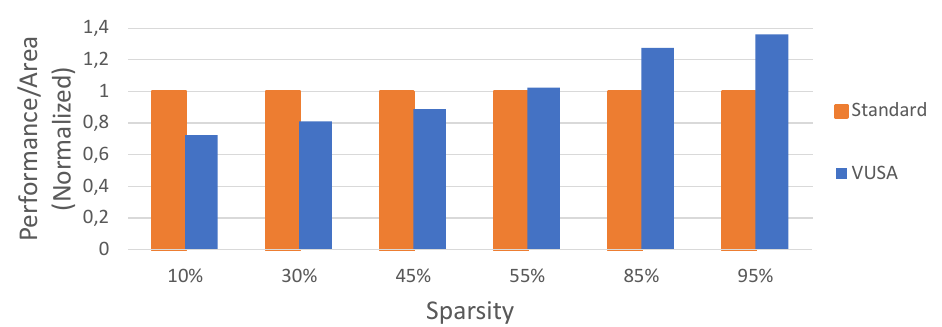}
    \squeezeCaption
    \caption{Area efficiency at different pruning rates.}
    \label{fig:6_eff_vs_prune_area}
    \vspace{1mm}
\end{figure}

The obtained results for the area efficiency are shown in Figure~\ref{fig:6_eff_vs_prune_area}. 
For a model pruned at $95\%$, there is a $36\%$ improvement for the VUSA compared to a standard $3\times 6$ systolic array, whereas for a base model without any pruning, there is a reduction of $28\%$. For a model pruned at $55\%$, the area efficiency is slightly higher for the VUSA. 

\begin{figure}[]
    \centering
    \includegraphics[trim=330 220 15 240,clip,width=0.48\textwidth]{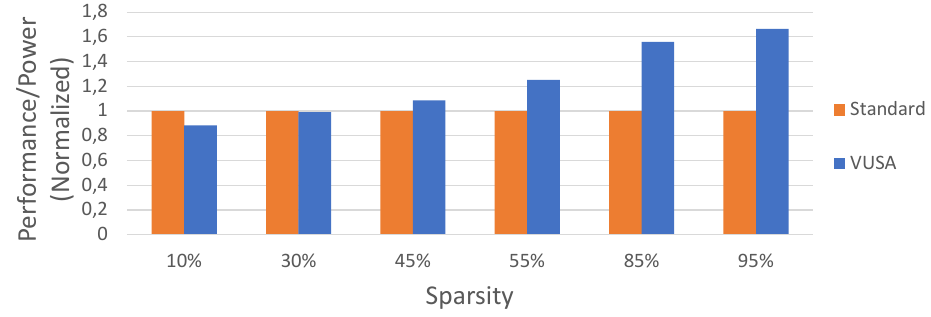}
    \squeezeCaption
    \caption{Power efficiency at different pruning rates.}
    \label{fig:6_eff_vs_prune_power}
    \squeezeText
\end{figure}

The power efficiency is shown in Figure~\ref{fig:6_eff_vs_prune_power}.
For a $95\%$ pruning rate, there is a $67\%$ improvement for the VUSA compared to a standard $3\times 6$ systolic array. For a base model without any pruning, there is a degradation by $11\%$. For a model pruned at $30\%$, the power efficiency is slightly lower for the VUSA. 

In summary, from a $30\%$ pruning rate, there are gains in power efficiency. At a $55\%$ pruning rate, there are gains in area efficiency as well. The stronger the pruning, the bigger the gains. It can also be observed that the VUSA exhibits great potential to improve power efficiency by 68\%.

\section{Related Work}
\label{sec:rw}

In several commercial and academic AI accelerators, systolic arrays are used to accelerate matrix multiplications which account for the largest fraction of the compute in DNN applications~\cite{tpu,why_systolic}. In the traditional systolic array, sparsity (i.e., zero-valued parameter or data) cannot be exploited.
Several extensions to systolic arrays have been proposed that allow to exploit sparsity if it follows pre-defined structures (e.g.,~\cite{sense,s2ta}). The drawback is a reduction in the maximum achievable sparsity and/or accuracy. 

Other published work aims at exploiting unstructured sparsity (e.g., \cite{eie,cnvlutin,eyeriss,scnn}), on either activations or weights like the technique presented in this paper. These reference architectures all introduce an area overhead to exploit unstructured weight sparsity. In contrast, our proposed VUSA achieves area savings by reducing the number of physical MAC units required for performing computations and introduces a simplified PE that processes sparse weights. Moreover, our proposed technique preserves the general scalability and programming model of the widely spread systolic array, which enables compatibility with existing tools and frameworks.

\section{Conclusion}
In this work, we introduced VUSA, a novel systolic array structure that can exploit unstructured sparsity by virtually growing to a larger size with the same number of physical MAC units. Experimental results for a commercial 16-nm technology and real-life DNN application show that our technique can improve the area and power efficiencies by around 55\% and 25\%, respectively, while additionally improving performance by 10\% compared to the baseline systolic array architecture. 
Still, the proposed architecture supports any type of DNNs, even without any sparsity.

\section*{Acknowledgments}
This research is funded by the German Federal Ministry of Education and Research within the project ”GreenEdge-FuE“, funding no. 16ME0517K.

This result is part of the IPCEI ME/CT and is funded by the European Union Next Generation EU, the German Federal Ministry for Economic Affairs and Climate Action, the Bavarian Ministry of Economic Affairs, Regional Development and Energy, the Free State of Saxony with the help of tax revenue based on the budget approved by the Saxon State parliament and the Free and Hanseatic City of Hamburg.

\bibliography{references}

\begin{thebibliography}{10}
\providecommand{\url}[1]{#1}
\csname url@samestyle\endcsname
\providecommand{\newblock}{\relax}
\providecommand{\bibinfo}[2]{#2}
\providecommand{\BIBentrySTDinterwordspacing}{\spaceskip=0pt\relax}
\providecommand{\BIBentryALTinterwordstretchfactor}{4}
\providecommand{\BIBentryALTinterwordspacing}{\spaceskip=\fontdimen2\font plus
\BIBentryALTinterwordstretchfactor\fontdimen3\font minus \fontdimen4\font\relax}
\providecommand{\BIBforeignlanguage}[2]{{%
\expandafter\ifx\csname l@#1\endcsname\relax
\typeout{** WARNING: IEEEtran.bst: No hyphenation pattern has been}%
\typeout{** loaded for the language `#1'. Using the pattern for}%
\typeout{** the default language instead.}%
\else
\language=\csname l@#1\endcsname
\fi
#2}}
\providecommand{\BIBdecl}{\relax}
\BIBdecl

\bibitem{cnn}
Z.~Li, F.~Liu, W.~Yang, S.~Peng, and J.~Zhou, ``A survey of convolutional neural networks: analysis, applications, and prospects,'' \emph{IEEE transactions on neural networks and learning systems}, 2021.

\bibitem{tpu}
N.~P. Jouppi \emph{et~al.}, ``In-datacenter performance analysis of a tensor processing unit,'' in \emph{Proceedings of the 44th annual international symposium on computer architecture}, 2017, pp. 1--12.

\bibitem{spots}
M.~Soltaniyeh, R.~P. Martin, and S.~Nagarakatte, ``An accelerator for sparse convolutional neural networks leveraging systolic general matrix-matrix multiplication,'' \emph{ACM Transactions on Architecture and Code Optimization (TACO)}, vol.~19, no.~3, pp. 1--26, 2022.

\bibitem{sense}
W.~Sun \emph{et~al.}, ``Sense: Model-hardware codesign for accelerating sparse cnns on systolic arrays,'' \emph{IEEE Transactions on Very Large Scale Integration (VLSI) Systems}, vol.~31, no.~4, pp. 470--483, 2023.

\bibitem{deepcompression}
S.~Han, H.~Mao, and W.~J. Dally, ``Deep compression: Compressing deep neural networks with pruning, trained quantization and huffman coding,'' \emph{arXiv preprint arXiv:1510.00149}, 2015.

\bibitem{eie}
S.~Han \emph{et~al.}, ``Eie: Efficient inference engine on compressed deep neural network,'' \emph{ACM SIGARCH Computer Architecture News}, vol.~44, no.~3, pp. 243--254, 2016.

\bibitem{bamberg2023synapse}
L.~Bamberg, A.~Pourtaherian, L.~Waeijen, A.~Chahar, and O.~Moreira, ``Synapse compression for event-based convolutional-neural-network accelerators,'' \emph{IEEE Transactions on Parallel and Distributed Systems}, vol.~34, no.~4, pp. 1227--1240, 2023.

\bibitem{bamberg2024exploiting}
L.~Bamberg, A.~Najafi, and A.~Garcia-Ortiz, ``Exploiting neural-network statistics for low-power {DNN} inference,'' \emph{IEEE Open Journal of Circuits and Systems}, 2024.

\bibitem{why_systolic}
H.-T. Kung, ``Why systolic architectures?'' \emph{IEEE computer}, vol.~15, no.~1, pp. 37--46, 1982.

\bibitem{scalesim}
A.~Samajdar, Y.~Zhu, P.~Whatmough, M.~Mattina, and T.~Krishna, ``Scale-sim: Systolic cnn accelerator simulator,'' \emph{arXiv preprint arXiv:1811.02883}, 2018.

\bibitem{cnvlutin}
J.~Albericio, P.~Judd, T.~Hetherington, T.~Aamodt, N.~E. Jerger, and A.~Moshovos, ``Cnvlutin: Ineffectual-neuron-free deep neural network computing,'' \emph{ACM SIGARCH Computer Architecture News}, vol.~44, no.~3, pp. 1--13, 2016.

\bibitem{eyeriss}
Y.-H. Chen, T.~Krishna, J.~S. Emer, and V.~Sze, ``Eyeriss: An energy-efficient reconfigurable accelerator for deep convolutional neural networks,'' \emph{IEEE journal of solid-state circuits}, vol.~52, no.~1, pp. 127--138, 2016.

\bibitem{sparsezoo}
\BIBentryALTinterwordspacing
NeuralMagic, ``{SparseZoo}: Neural network model repository for highly sparse and sparse-quantized models with matching sparsification recipes,'' accessed: 2023-12-21. [Online]. Available: \url{https://sparsezoo.neuralmagic.com}
\BIBentrySTDinterwordspacing

\bibitem{resnet}
K.~He, X.~Zhang, S.~Ren, and J.~Sun, ``Deep residual learning for image recognition,'' in \emph{Proceedings of the IEEE conference on computer vision and pattern recognition}, 2016, pp. 770--778.

\bibitem{mobilenet}
A.~G. Howard \emph{et~al.}, ``Mobilenets: Efficient convolutional neural networks for mobile vision applications,'' \emph{arXiv preprint arXiv:1704.04861}, 2017.

\bibitem{s2ta}
Z.-G. Liu, P.~N. Whatmough, Y.~Zhu, and M.~Mattina, ``S2ta: Exploiting structured sparsity for energy-efficient mobile cnn acceleration,'' in \emph{2022 IEEE International Symposium on High-Performance Computer Architecture (HPCA)}.\hskip 1em plus 0.5em minus 0.4em\relax IEEE, 2022, pp. 573--586.

\bibitem{scnn}
A.~Parashar, M.~Rhu, A.~Mukkara, A.~Puglielli, R.~Venkatesan, B.~Khailany, J.~Emer, S.~W. Keckler, and W.~J. Dally, ``Scnn: An accelerator for compressed-sparse convolutional neural networks,'' \emph{ACM SIGARCH computer architecture news}, vol.~45, no.~2, pp. 27--40, 2017.

\end{thebibliography}
\bibliographystyle{IEEEtran}

\end{document}